# Mobile-Based COVID-19 Vaccination Registration Application Prototype

**Leon A. Abdillah** [1)*], **Azka Kurniasti** [2)]
[1, 2)] Department of Information Systems, Universitas Bina Darma, Indonesia

[1)] leon.abdillah@yahoo.com

**Abstract:** Information technology-based applications have entered the era of mobile phones or smartphones such as those using the Android or iOS operating system. Mobile-based application development has become a trend for today's society. Especially during the global COVID-19 pandemic, almost all activities are carried out remotely through mobile-based applications. To prevent the spread of COVID-19, mass vaccines are given to the public. So that the process of administering the vaccine does not cause crowds, it is necessary to create a mobile-based application. So that the application can be further developed properly, it is necessary to make a prototype. The prototype consists of 5 (steps): 1) Quick plan, 2) Modeling Quick Design, 3) Construction of prototype, 4) Deployment Delivery & feedback, and 5) Communication. In this research, the InVision design tool is used which can help design prototypes for both mobile and web versions. InVision has been widely used in making prototypes and is used by many digital companies in the world. The results obtained are in the form of a prototype application for the registration of vaccine participants via mobile phones and also the web. The programmers will easily translate the prototype results into a mobile-based application for the benefit of mobile phone-based online vaccine registration.

## INTRODUCTION

Information and Communication Technology (ICT) has proven to be an icon of progress in the 21st century. ICT has become a node for all aspects of life in the era of globalization with the trend of digitization. ICT itself follows the trend of globalization so that it can be enjoyed throughout the world equally (Abdillah, Handayani, Rosalyn, & Mukti, 2021). There are at least 5 (five) IT Megatrends (Valacich & Schneider, 2018)that are currently dominant, namely: 1) Big Data, 2) Cloud Computing, 3) IoT, 4) Smartphones, and 5) Social Media. The use of gadgets in the form of smartphones makes it easier for all people's activities. Mobile devices (Sharp, Rogers, & Preece, 2019) have become increasingly widespread, with more and more people using them in various aspects of their daily lives and in their work. With increasingly sophisticated smartphones, many information technology-based services have emerged. One of the leading smartphone operating systems that are widely used is products from Apple that use the iOS operating system for their Apple smartphone gadgets.

In July 2020, Apple was listed as the most valuable brand in the world (Alsop, 2020). Even at the beginning of 2020, Apple products that are active around the world have sold over 1 billion gadgets. Since its founding in 1976 by Steve Jobs, Apple has continued to be a major player in the gadget and smartphone industry. Even though they are offered at a high price, Apple products are still selling well in the market and the Apple brand is still flying in the ranks of the world's technology brands. Along with the increasing number of Apple smartphone users based on iOS operating systems. So, more and more mobile applications (Abdillah et al., 2020) are being developed for Apple's smartphone products. Figure 1 shows Apple's position as a top brand surpassing Microsoft, Google, Tencent, Facebook, IBM, etc.

Software that supports the development of the prototype is available for free in cloud mode. One of the most widely used cloud-based prototype software is InVision. Besides being available for free in the cloud, InVision also provides InVision Studio for designers who need it.

Globally, the total number of COVID-19 cases in Indonesia ranks 15[th] (Table 1). However, for new cases reported in the last 24 hours, Indonesia ranks 6th. With the very high total number of new cases and cases in Indonesia, both the government and the private sector are working hard to prevent the spread of COVID-19 transmission in Indonesia.

Until mid-2021, no cure for COVID-19 has been found. To reduce the impact of COVID-19, a number of agencies from several countries have made vaccines. Vaccines are given massively to reduce the negative impact of exposure to COVID-19. There are several types of COVID-19 vaccines, including: 1) AstraZeneca (UK), 2) Moderna (USA), 3) Pfizer (USA), 4) Novavax (USA), 5) Sinopharm (China), 6) Sinovac (China), and 7) Bio Farma (Indonesia).

---

*Corresponding author: Leon A. Abdillah (leon.abdillah@yahoo.com)

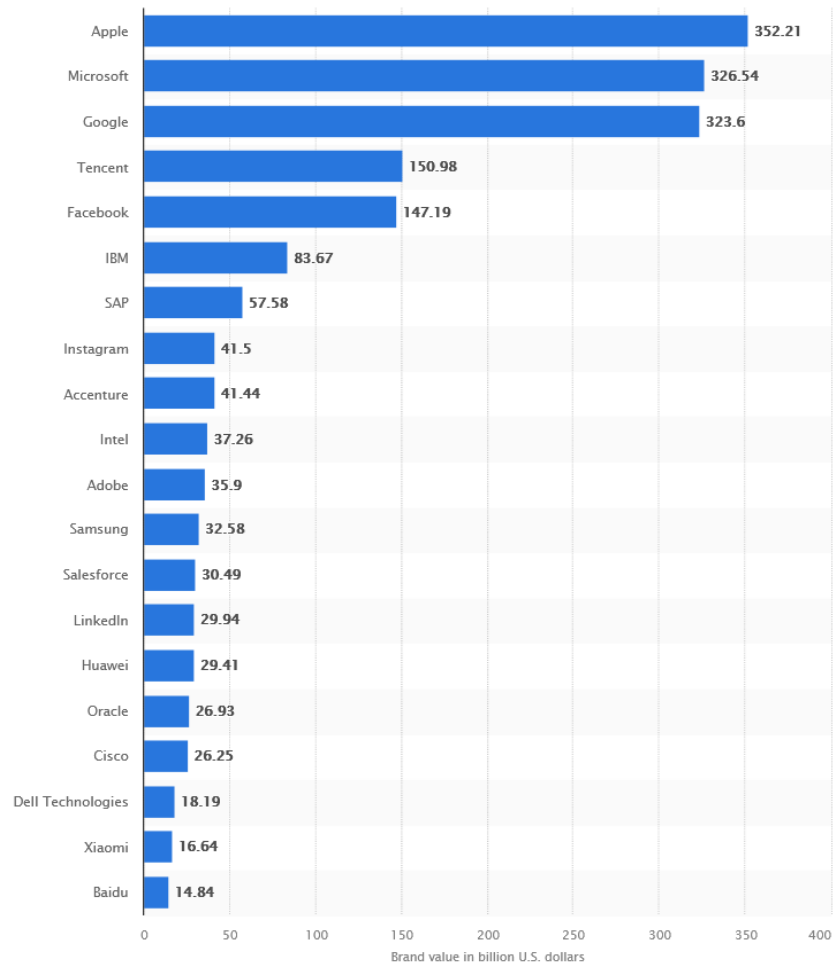

Fig. 1 World Top Brand Prototype Development (Alsop, 2020).

In a pandemic where the cure for a new disease has not been found, the vaccine is the best solution. Vaccines (Harrison & Wu, 2020) play a key role in reducing the impact of the disease on the mortality rate it causes. The COVID-19 vaccination aims to reduce the transmission/transmission of COVID-19, reduce morbidity and mortality due to COVID-19, achieve herd immunity, and protect the community from COVID-19 in order to remain socially and economically productive (Kemenkes RI, 2021). The pandemic of COVID-19 has forced every party to use and expand IT-based and digital services (Abdillah, Mukti, Puspita, & Suhartini, 2021).

Table 1. Top 15 countries, areas or territories of COVID-19 cases (13 Aug 2021) (WHO, 2021).

| Rank | Country | Continent | Cases |
| --- | --- | --- | --- |
| 1 | USA | America | 36,099,344 |
| 2 | India | Asia | 32,117,826 |
| 3 | Brazil | America | 20,245,085 |
| 4 | Russian | Europe/Asia | 6,557,068 |
| 5 | France | Europe | 6,244,939 |
| 6 | UK | Europe | 6,179,510 |
| 7 | Turkey | Europe/Asia | 6,018,485 |
| 8 | Argentina | America | 5,052,884 |
| 9 | Colombia | America | 4,852,323 |
| 10 | Spain | Europe | 4,677,883 |
| 11 | Italy | Europe | 4,420,429 |
| 12 | Iran | Asia | 4,320,266 |

*Corresponding author: Leon A. Abdillah (leon.abdillah@yahoo.com)

2153

The rest of this paper contains a literature review (Section II), followed by research procedures (Section III), then describes the results and discussion (Section IV), and closes with conclusions (Section V).

## LITERATURE REVIEW

A number of previous studies are used as references, such as 1) Making User-focused Prototype: Using Design Sprint to Test, Design, and Prototype Mobile App Rapidly (Wong, 2016), 2) Fit Buddy Prototype and KSUGo Mobile App (Lim, 2018), and 3) Designing Ajri Learning Journal Center Using Invision Tools to Realize Creative Innovation Soft Skills (Hariguna & Wahyuningsih, 2020). Of these studies, no one has made a prototype for the COVID-19 vaccine registration process.

Comparisons were made with a number of relatively equal studies (Table 2). This study creates a prototype that can be developed for the latest mobile-based applications, uses the prototype method, and can load vaccine variants that suit the needs of COVID-19 vaccination.

Table 2. The Comparison with Similar Studies.

| No | Similar Studies | Methods | Platforms |
| --- | --- | --- | --- |
| 1 | Sistem Informasi Pendaftaran Vaksinasi Covid-19 (Nurhadi & Indrayuni, 2021) | Waterfall | Website |
| 2 | Pemanfaatan Aplikasi Google Form dalam Kegiatan Serbuan Vaksinasi COVID-19 Berbasis Online di Kabupaten Klaten (Wicaksono & Setiawan, 2022) | Participatory Rural Appraisal (PRA) | Google Forms |
| 3 | Perancangan Sistem Informasi Pendaftaran Online Vaksinasi COVID-19 di RSAU Dr. M. Salamun (Tarigan & Mahpud, 2021) | Object Oriented Programming (OOP) | Website |

## METHOD

The third part of this paper presents 3 (three) aspects related to the implementation of research, which consist of: 1) Research Procedures, 2) Prototype, 3) Research Tools.

**Research Procedure**

On-site vaccination process (Figure 2): 1) Come according to the schedule that has been registered, 2) Registration to check whether prospective vaccine recipients are registered or not, and the data verification process in the vaccination system, 3) Health checks in the form of checking blood pressure & temperature before vaccination, and a screening process is carried out, 4) If you pass the health examination stage, the vaccination process is carried out, 5) Observation for 30 minutes, and 6) Vaccination is complete and the proof is printed that the person concerned has been vaccinated.

In this study, the focus is on the registration process to get the vaccine, which is carried out through an iOS-based mobile application. Documents that need to be prepared include 1) ID card, 2) active email, 3) active cellphone number, 4) invitation letter or notification of vaccine schedule if any.

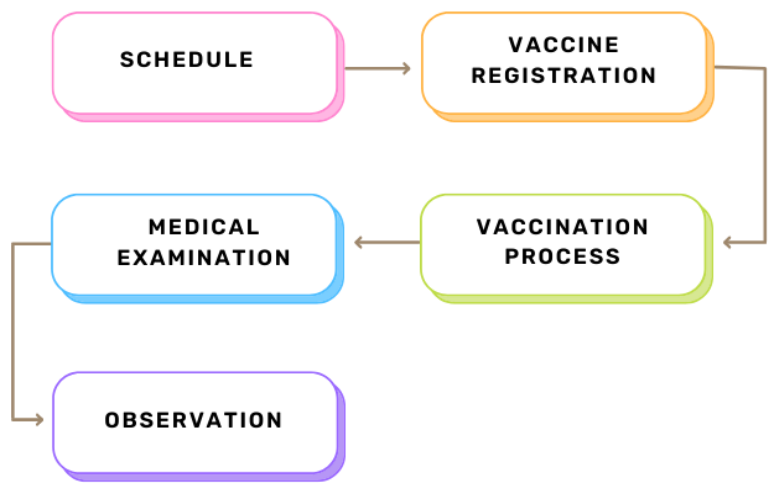

Fig. 2 Vaccination Procedure Flowchart.

*Corresponding author: Leon A. Abdillah (leon.abdillah@yahoo.com)

**Prototype**

The prototype is one form of software development (Abdillah, Atika, Kurniawan, & Purwaningtias, 2019) and a powerful design tool (Shneiderman et al., 2018). The prototype approach is widely used to cover situations where there are no detailed functions, features, algorithms, etc. The prototype paradigm (Pressman & Maxim, 2020) consists of 5 (five) stages (Figure 3): 1) Communication, 2) Quick Plan, 3) Modeling Quick design, 4) Construction of prototype, and 5) Deployment Delivery & feedback. Rapid prototypes can be better in speed, accuracy, and less laborious than conventional methods of creating models (Neeley, Lim, Zhu, & Yang, 2013).

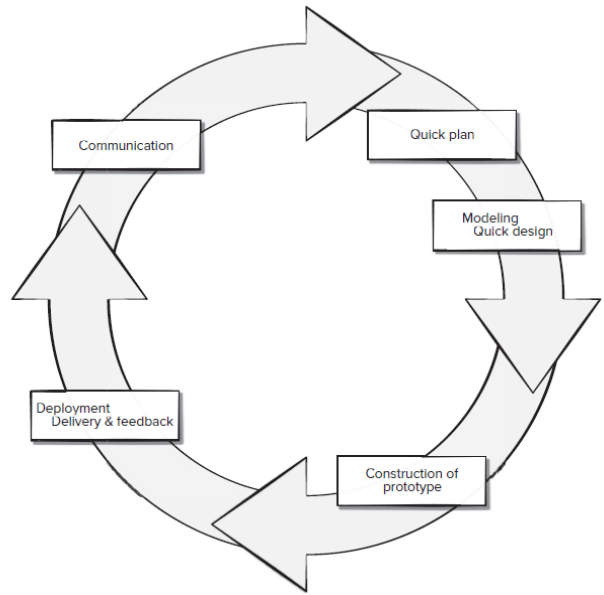

Fig. 3 The Prototype Paradigm (Pressman & Maxim, 2020)

**Research Tools**

InVision is a tool that is very convenient to use to make prototypes of various application designs, both web-based and mobile-based applications. With InVision, designers do not need to install it because InVision can be accessed online or cloud-based. InVision can be used for digital design, workflow tools, media collaboration, as well as prototyping tools (Lim, 2018). In an interview published in June 2021, InVision (Santiago, 2021) topped the list of "Most Popular Prototyping Tools" ahead of Adobe XD, Figma, Sketch, and Framer.

**RESULT**

After the prototype is completed, testing is carried out on the links between modules. After testing all the links are running and connected to the appropriate nodes. The result of this research is a prototype that serves the vaccine registration process for people who will carry out the COVID-19 vaccine. The results and discussion will be divided into 8 (eight) sections, namely: 1) Home Screen, 2) Registration, 3) Verification, 4) Registration Successful, 5) Admin Dashboard Menu, 6) Admin Menu "Vaccine Registrasi", 7) Admin Vaccine Data Menu, and 8) Admin Edit Menu.

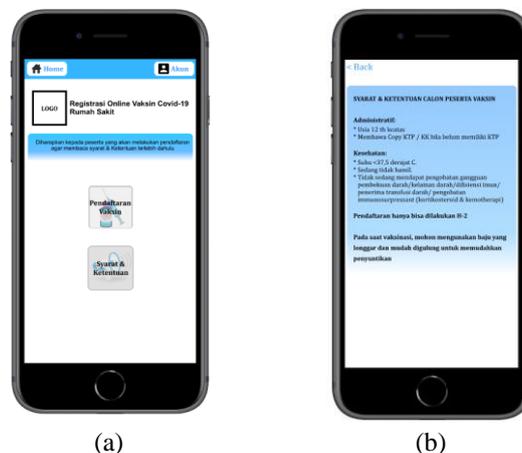

(a)          (b)
Fig. 4 Home Screen Display

*Corresponding author: Leon A. Abdillah (leon.abdillah@yahoo.com)

The Home Screen is the part of the application that first appears or is seen by the user. The home page shows the name and logo of the hospital, then the display of vaccine registration and conditions. In the header, there is a "Home" and "Akun" menu (Figure 4. a). If the "Syarat & Ketentuan" menu is clicked, the requirements for COVID-19 vaccination will appear (Figure 4. b).

## DISCUSSIONS

In this section, the researchers can give a simple discussion related to the results of the research trials. This section contains the author's opinion about the research results obtained. Common features of the discussion section include the comparison between measured and modeled data or comparison among various modeling methods, the results obtained to solve a specific engineering or scientific problem, and further explanation of new and significant findings

**Registration**

The registration menu is a menu used for the registration of new users of the application. To register, users can click the "Pendaftaran Vaksin" menu on the home page (Figure 4. a). Next, it will go to "Cek Validasi NIK". Users are asked to enter their NIK and Full Name, and then click the "Cek" button (Figure 5. a). "Cek Validasi NIK" is a feature to check whether the user is already registered in the application.

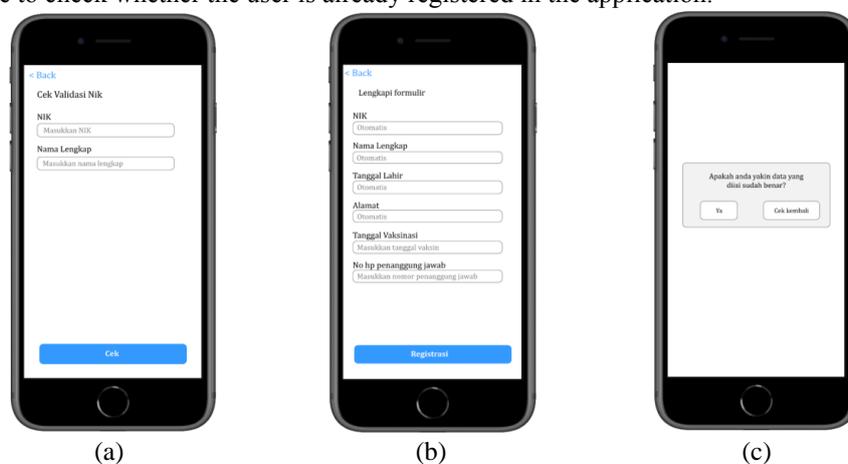

(a)     (b)     (c)

Fig. 5 Registration

If user data does not exist, the user is asked to complete the form by entering the NIK data, Full Name, Date of Birth, Address, Vaccination Date, and Mobile Number of the Person in Charge, and then click the "Registrasi" button (Figure 5. b). After that, a confirmation dialog box appears. If still in doubt, the user can click the "Cek Kembali" button (Fig. 5.c).

**Verification**

If users are sure the data entered is correct then click the "Yes" button, then a verification page will appear for user authentication (Figure 6. a). Users can choose to send a one-time password (OTP) code either via short message service (SMS) or Electronic mail (email or e-mail). An example of a verification page via SMS can be seen in Figure 6. b. After entering the 6-digit OTP code then press the "Register" button.

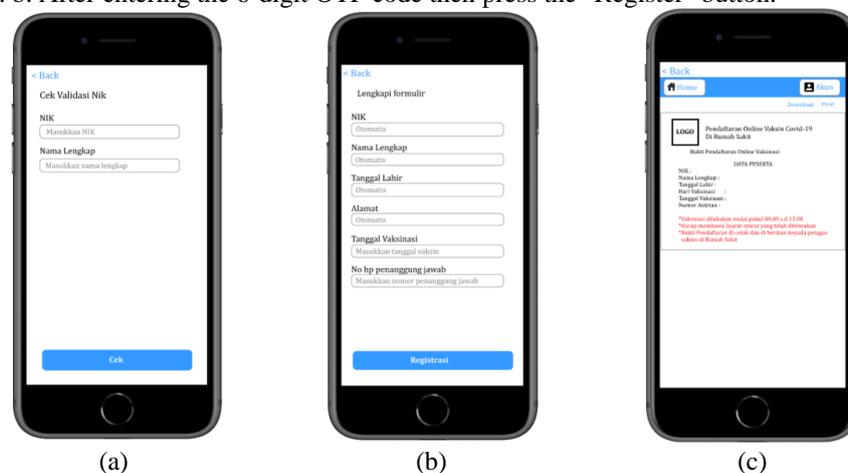

(a)     (b)     (c)

Fig. 6 Registration

*Corresponding author: Leon A. Abdillah (leon.abdillah@yahoo.com)

One-Time Passwords (OTP) is an added security feature. OTP behaves as a User ID and Password representative for a certain period of time. OTP provides additional protection from two-factor authentication mechanisms for more secure applications (Huang, Huang, Zhao, & Lai, 2013). The OTP is only generated for one-time use. If the provided period of time has passed, it is necessary to generate a new OTP again. By applying the OTP scheme (Tzemos, Fournaris, & Sklavos, 2016), the security authentication of application users will be better or increased.

After the user enters the OTP code in the form of a 6 (six) digit number correctly, the registration is successful and the "Bukti Pendaftaran Online Vaksinasi" page will appear (see Figure 6.c). Evidence of online vaccination registration contains information on "Participant Data" to be vaccinated, in the form of 1) NIK, 2) Full Name, 3) Date of Birth, 4) Vaccination Day, 5) Vaccination Date and 6) Queue Number. At the bottom, it is also informed about the time of vaccination, vaccination requirements, and proof of registration.

**Admin Menu Dashboard**

This prototype is also equipped with an Admin page. Admin can perform a number of actions, such as 1) Login, 2) Dashboard (Figure 7), 3) Vaccine Registrant Data (Figure 8), 4) Vaccination Data (Figure 9), and 5) Edit Vaccination Data (Figure 10). After the Admin has successfully logged in, it will enter the "Dashboard" page. This page displays a menu on the left side, then in the middle, it is equipped with an image of "Grafik Pendaftar" (Figure 8). The "Logout" button is at the top right.

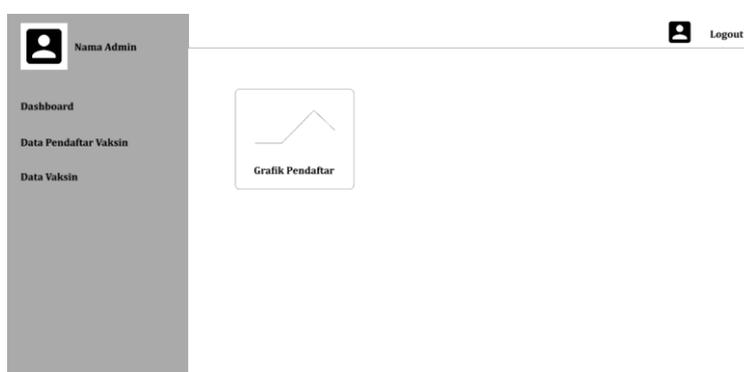

Fig. 7 Admin Dashboard.

**Admin Menu "Pendaftar Vaksin"**

The "Data Pendaftar Vaksin" menu contains a search facility based on NIK or Date. NIK is entered through the text field object, while the Date is entered through the Date & Time object. The list of participants will be displayed in a table consisting of 4 (four) columns, namely: 1) NIK, 2) Name, 3) Vaccine Schedule, and 4) Action. The search results can also be downloaded or printed in PDF (see Figure 8).

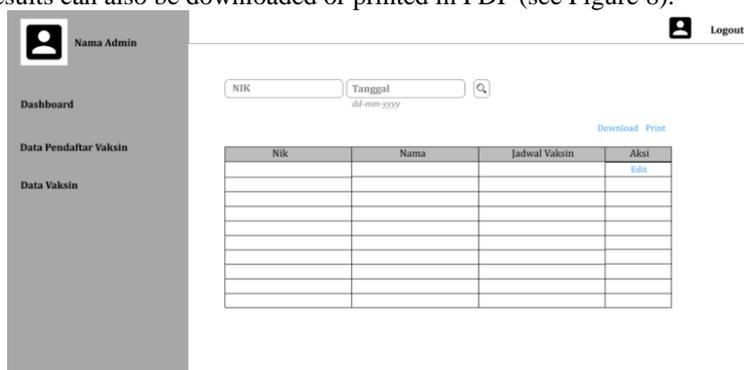

Fig. 8 Menu of "Data Pendaftar Vaksin".

**Admin Menu "Data Vaksin"**

The "Vaccine Data" menu contains a list of participants who have been vaccinated. At the top there are 2 (two) buttons, namely: 1) "Vaksin Ke-1" button, and 2) "Vaksin Ke-2" button. There is a search facility based on NIK. The results will be displayed in the form of a table consisting of 5 (five) columns, namely: 1) NIK, 2) Name, 3) 1st Vaccine, 4) 2nd Vaccine, and 5) Action. See Figure 9 for this example. The data search is done by

---

entering the NIK through the text field object, then the search process is carried out when the user clicks the "Search" button there is a search icon.

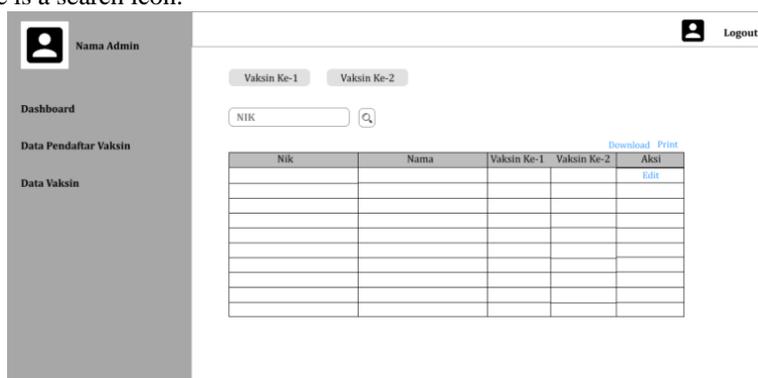

Fig. 9 Menu of "Data Vaksin".

**Admin Edit Menu**

Admin is also equipped with facilities to correct data based on NIK. These improvements can be in the form of Name changes, 1st Vaccine information, and 2nd Vaccine Information, as well as "Kembali" and "Simpan" buttons (see Figure 10). NIK and Name will be entered through the text field object, while the information about the vaccine will be entered through the radio button object.

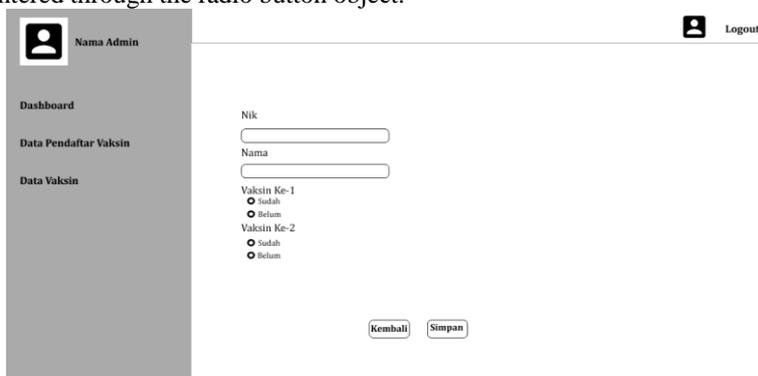

Fig. 10 Edit Menu.

## CONCLUSION

The COVID-19 vaccine patient registration prototype makes it easy for programmers to develop iOS-based mobile applications. This prototype can be used not only for iOS Apps but also compatible or can be used for the development of Android Apps or other mobile applications. For further research, a mobile app can be built for the COVID-19 vaccine registration process. Last but not least, the Author also intends to extend the research involving knowledge management systems (Abdillah, Sari, & Indriani, 2018) concepts.

*Corresponding author: Leon A. Abdillah (leon.abdillah@yahoo.com)

*Corresponding author: Leon A. Abdillah (leon.abdillah@yahoo.com)